\documentclass[a4paper]{jpconf}
\usepackage{graphicx}
\usepackage{adjustbox}
\usepackage{multicol}
\usepackage{gensymb}
\usepackage{subfig}

\begin{document}
\title{Probing dark matter in 2HDM+S with MeerKAT Galaxy Cluster Legacy Survey data}

\author{N Lavis and G Beck}

\address{School of Physics, University of the Witwatersrand,Private Bag 3, WITS-2050, Johannesburg,South Africa}

\ead{1603551@students.wits.ac.za, geoffrey.beck@wits.ac.za}

\begin{abstract}
	Dark matter is believed to constitute the majority of the matter content of the universe, but virtually nothing is known about its nature. Physical properties of a candidate particle can be probed via indirect detection by observing the decay and/or annihilation products. While this has previously been done primarily through gamma-ray studies, the increased sensitivity of new radio interferometers means that searches via the radio bandwidth are the new frontrunners. MeerKAT's high sensitivity, ranging from 3 $\mu$Jy beam$^{-1} $ for an 8 arcsecond beam to 10 $\mu$Jy beam$^{-1} $ for an 15 arcsecond beam, make it a prime candidate for radio dark matter searches. Using MeerKAT Galaxy Cluster Legacy Survey (MGCLS) data to obtain diffuse synchrotron emission within galaxy clusters, we are able to probe the properties of a dark matter model. In this work we consider both generic WIMP annihilation channels as well as the 2HDM+S model. The latter was developed to explain various anomalies observed in Large Hadron Collider (LHC) data from runs 1 and 2. The use of public MeerKAT data allows us to present the first WIMP dark matter constraints produced using this instrument.
\end{abstract}

	\section{Introduction}
The unknown nature of dark matter remains an unanswered question within our current cosmological model. Various methods have been employed as probes namely indirect searches of annihilation/ decay products, collider searches or direct detection.
The two-Higgs doublet model with an additional singlet scaler (2HDM+S) is a particle physics model containing a dark matter candidate. The model has been put forward as an explanation to various anomalies observed in Large Hadron Collider data from runs 1 and 2 \cite{von2015compatibility,von2016phenomenological}. It is of particular interest as the proposed mass of the dark matter candidate overlaps with various astrophysical models that aim to explain the anti-particle and gamma-ray excesses seen at the galactic centre by Fermi-LAT \cite{Fermi-LAT} and HESS \cite{HESSsite}. Previous indirect dark matter searches have primarily been performed using gamma-ray experiments, such as Fermi-LAT \cite{atwood2009large} and HESS \cite{HESSsite}, due to the low attenuation and high detection efficiency they are capable of producing.
Radio frequency dark matter searches are becoming more prevalent due to radio interferometers having the advantage of superior angular resolution. This advantage  limits confusion over diffuse emission produced via dark matter annihilation and unresolved point sources.By modelling the expected synchrotron flux produced as a consequence of the dark matter candidate's annihilation within the astrophysical environment of galaxy clusters and comparing to the MeerKAT Galaxy Cluster Legacy Survey (MGCLS) data \cite{knowles2022meerkat}, we are able to produce limits on the annihilation cross sections for various masses and decay channels. These proceedings are structured as follows: Section 2 briefly discusses the 2HDM+S formalism and the dark matter sector it introduces. Section 3 introduces the synchrotron emission formalism with which the simulations are performed. Section 4 discusses MeerKAT data products and the methodology used to extract diffuse fluxes. Section 5 presents the results  \\
\section {2HDM+S and Dark Matter }
Following the discovery of the Higgs boson \cite{higgs1964broken,higgs1964broken1, englert1964broken,guralnik1964global} various multi-lepton anomalies have been observed in run 1 and 2 data at the LHC (\cite{von2015compatibility} \cite{von2016phenomenological}). An analysis of the multi-lepton final states deviate from the Standard Model (SM) predictions, hinting at physics beyond SM (BSM). An implication of the  2HDM+S model is the production of multiple leptons via its decay chain  H $\rightarrow$ Sh, SS \cite{von2016phenomenological}. The masses of the scalars H and S were fixed to  $m_H = 270$ GeV and $m_S =$ 150 GeV respectively \cite{von2018multi}. Statistically compelling excesses in various final states ( opposite and same sign di-leptons, and the three lepton channel with and without the presence of b-tagged jets) have been reported in references \cite{von2019emergence} \cite{von2020anatomy} \cite{hernandez2021anomalous}. In addition evidence for the production of the scalar S with mass 151 GeV was obtained by combining side band data from SM Higgs searches \cite{crivellin2021accumulating}. When all decay channels are included a global significance of 4.8 $\sigma$ was reported for the required mass range (130 -160 GeV) to explain the anomalies \cite{crivellin2021accumulating}. The scalar S can potentially act as a mediator between SM particles and the dark matter candidate introduced within the hidden sector of the model. 

\section{Synchrotron emission model}

The formalism for predicting the surface brightness of synchrotron emission within a given halo environment is outlined by Beck et al in \cite{beck2019radio}.  
The power of synchrotron emission produced by an electron of energy E within a magnetic field of strength B is given by \cite{longair2010high} as: 

\begin{equation}
P_\mathrm{sync}(\nu,E,r,z) = \int_{0}^{\pi} d\theta \frac{\sin ^2\theta}{2} 2\pi \sqrt{3} r_e me_e c \nu_g F_{sync} (\frac{\kappa}{\sin\theta})
\end{equation}

where $\nu$ is the observed frequency, z is the redshift of the source, $m_e$ is the mass of an electron, $\nu_g= \frac{c B}{2 \pi m_e c} $ is the non-relativistic gyro-frequency and $r_e = \frac{e^2}{m_e c^2}$ is the classical radius of an electron \cite{beck2019radio}.

The parameter $\kappa$ is defined as 
\begin{equation} 
\kappa = \frac{2 \nu (1+z)}{3\nu_0 \gamma} \left( 1+ \left(\frac{\nu_p \gamma}{\nu(1+z)} \right)^2  \right)^{3/2}
\end{equation}

where $ \nu_p $ is the plasma frequency, which is directly dependent on the electron density of the environment. The parameter $F_{sync} $ describes the synchrotron kernel and is defined as 

\begin{equation}
F_{sync} (x) = x \int_{x}^{\infty} dy K_{5/3}(y) \approx 1.25 x^{1/3} e^{-x} (648 +x^2)^{1/12}
\end{equation}

The synchrotron emissivity at a radial position r within a halo is then found to be 
\begin{equation}
j_{sync} (\nu,r,z) = \int_{m_e}^{M_\chi} dE \left( \frac{dn_{e^-}}{dE} + \frac{dn_{e^+
}}{dE}\right)P_{sync}(\nu,E,r,z)
\end{equation}
The factor $\frac{dn_e}{dE}$ describes the particle (electron and positron respectively) equilibrium distribution. When considering dark matter induced radio emission the diffusion and energy loss experienced by the  resultant electrons must be considered. This is due to the fact that position and energy distributions of the electrons will influence the subsequent synchrotron emission \cite{beck2019radio}. The equilibrium distributions can be found by solving the diffusion equation for a dark matter halo 

\begin{equation}
\frac{\partial}{\partial t} \frac{dn_e}{dE} = \nabla \left(D(E, \mathbf x) \nabla \frac{dn_e}{dE} \right) + \frac{\partial}{\partial E} \left(b(E, \mathbf x ) \frac{n_e}{dE} \right) + Q_e(E,\mathbf x)
\end{equation}

In the above equation $\frac{dn_e}{dE} $ is the electron spectrum, the spatial diffusion is described with $ D(E, \mathbf x)$, $b(E, \mathbf x )$ describes the rate of energy loss and the electron source function is given by the function $ Q_e(E,\mathbf x)$. Typical methods for solving the diffusion equation are outlined in \cite{beck2019radio}. 
The flux density spectrum within a radius r of the halo centre is then found to be 

\begin{equation}
S_{sync}(\nu ,z) = \int_{0}^{r} d^3 r' \frac{j_{sync}(\nu,r',z) }{4 \pi D_L^2}
\end{equation}
where $D_L$ is the luminosity distance to the source in question \cite{beck2019radio}.

\section{MGCLS}
Galaxy clusters are the largest gravitationally bound systems in the universe, with their composition dominated by dark matter. This makes them promising astrophysical labs to study potential signatures of dark matter. Radio observations of clusters have revealed steep-spectrum diffuse radio emission (see reviews  \cite{feretti2012clusters},\cite{van2019diffuse}). This data can potentially be used to probe the distributions of cosmic ray particles and cluster magnetic fields \cite{knowles2022meerkat}. Having more accurate representations of the magnetic field environments within the targets of interest may potentially reduce the uncertainties of the expected synchrotron  emission. This will in turn lead to potentially tighter constraints on dark matter properties. As found in this study and others \cite{storm2013constraints}, a limited number of clusters have well studied magnetic fields. Thus MeerKAT's potential for studying magnetic fields may greatly benefit attempts at constraining dark matter.
The MeerKAT telescope is currently in operation under the South African Radio Astronomy Observatory (SARAO), with 64 13.5 m-diameter antennas, creating a powerful instrument for wide area surveys, with high sensitivity over a range of angular scales \cite{knowles2022meerkat}.   
With sensitivity ranging from 3 $\mu$Jy beam$^{-1} $ for an 8 arcsecond beam to 10 $\mu$Jy beam$^{-1} $ for an 15 arcsecond beam \cite{knowles2022meerkat}.
MGCLS consists of approximately 1000 hours of observations in the L-band ( 900-1670 MHz) of 115 galaxy clusters in full polarization between -80$\degree$ and 0$\degree$ declination. The L-band system has a primary beam FWHM of 1.2$\degree$ at 1.28GHz \cite{knowles2022meerkat}.
The galaxy cluster sample has two sub-components, a radio selected sample and an X-ray selected sample. The radio selected sample consisting of 41 clusters that have previously been searched for diffuse radio emission. The X-ray selected clusters were chosen from the MCXC catalogue to fill gaps in the observation schedule as required.
For this analysis 7 clusters containing radio halos were considered, all with a data quality flag higher than 3 \cite{knowles2022meerkat}.
In order to study the diffuse emission the enhanced data products of MGCLS were utilized. These consist of 5 plane cubes in both full resolution (7.5-8'') and convolved (15'') resolution. The methodology of measuring the diffuse flux of radio halos is as follows:
SAODS9 radio flux plugin \cite{SAODS9} was used to measure the radio flux of the desired region. The plugin measures the flux within a specified region, and subtracts the background flux. This plugin also produces an estimated statistical error. This measurement is performed on the convolved image. In order to consider only the diffuse emission the fluxes due to compact sources must be removed. In order to do this the source finding method as in MGCLS is followed.   	
Python Blob Detector and Source Finder \cite{mohan2015pybdsf} is utilized. This package searches for islands of emission and attempts to fit models of elliptical Gaussians. These are then grouped into sources. The default settings were not varied, these are aof a $3 \sigma_\mathrm{image} $ island boundary and a $ 5\sigma_\mathrm{image} $ source detection threshold. Note that $\sigma_{image} $ is the local image RMS, and the program varies this across the field as required. In order to produce more accurate source catalogues this process was performed on the full resolution images of each cluster. \\
The sum of the compact fluxes is then removed from the total flux of the region. Estimation of the error in the diffuse flux is a sum of squares of the statistical error estimated by DS9 and a 5\% error for calibration and systematic effects as quoted in \cite{knowles2022meerkat}.   
\section{Results and Discussion}
Presented here are the results for seven galaxy clusters. This sample contains five radio selected clusters and two X-ray selected clusters. The physical properties required to simulate the halo environment of each cluster are listed in Table 1. \\
The predicted annihilation fluxes were produced under the assumption that the dark matter distribution is smooth. In physical halos, it is known that this is not the case. In the $\Lambda$CDM paradigm it is predicted that halos would contain self-bound structure, sub-halos. This is a direct consequence of bottom- up structure formation, where larger halos are formed through mergers and accretion of smaller ones. Since the dark matter annihilation signal is proportional to the square of the density it is expected that the presence of these more concentrated regions will have the effect of enhancing the signal \cite{sanchez2014flattening, moline2017characterization}. This enhancement, or boosting effect is of greater importance in more massive halos as they are expected to enclose more hierarchical levels of structure formation.
The total halo boost factors can be calculated using the parametric equation in \cite{moline2017characterization} with $\alpha$=2. For the masses of the clusters considered the total boost factor is approximately 60. However this factor is produced mainly for a $\gamma$-ray signal. Synchrotron emissions will not experience this full boost factor, as sub-halos are more common around the outskirts of the host halo. In these regions the magnetic fields are generally much weaker. Thus it is necessary to calculate a scaled boost factor. This can be accomplished by multiplying the distribution of the host halo with a modification function from \cite{jiang2017statistics} in order to obtain the mass distribution of the sub-halos. This density is then normalized to produce a probability distribution. The scaled boost factor is then the sub over the probability of a sub-halo being at the given radius multiplied by the magnetic power factor at that point. The calculated values for each cluster are listed in the table below. 
	\begin{table}[h]
	\centering
	\label{table:1} 
	\caption{ Physical characteristics of the clusters. Column 2: redshift. Column 3: virial mass. Column 4: halo scale radius- defined as the virial radius divided by the virial concentration. Common alternate names are provided in column 5 and the scaled boost factor is given in column 6.} 
	 
	\begin{tabular}{|c|c|c|c|c|c|c|}
		\hline 
		Cluster name & z & M$_\mathrm{vir}$ (10$^{15}$ M$_{\odot}$) & R$_{s}$ (Mpc)   &Alternate name &Scaled Boost& References\\ 
		\hline 
		Abell 209 & 0.206 & 1.349 & 0.6140844507 & &5.69 &  \cite{knowles2022meerkat} \cite{klein2019weak} \\
		\hline
		Abell 370 &0.375 & 3.03& 0.364   &  G172.98-53.55&5.78 & \cite{knowles2022meerkat} \cite{lee2020nature} \cite{bartelmann1996arcs} \\
		\hline 
		Abell 2813  & 0.292 &1.241 & 0.6086956522  &J0043.4-2037 &2.84 & \cite{knowles2022meerkat} \cite{klein2019weak}\\
		\hline 
		Abell S295 &0.3 &0.511 & 0.4432432432 & J0245.4-5302&5.58 &\cite{knowles2022meerkat} \cite{klein2019weak}\\
		\hline 
		Abell S1063 & 0.348 &1.489 &0.6636363636  &  J2248.7-4431 &5.74 & \cite{knowles2022meerkat} \cite{klein2019weak} \\
		\hline
		J0528.9-3927 & 0.284&1.64 &0.6525 &   &1.82 &\cite{knowles2022meerkat} \cite{foex2017core} \\
		\hline 
		J0645.4-5413 &0.167 &1.240 &0.6109589041   & Abell 3404&4.27 & \cite{knowles2022meerkat} \cite{klein2019weak} \\
		\hline  
	\end{tabular}
	 
\end{table}

	The mass range that has been considered is 75-200 GeV in order to overlap with the
mass range of the 2HDM+S dark matter candidate expected from kinematic considerations. \cite{von2016phenomenological}. The simulated fluxes are compared to the measured values with a $2\sigma$ confidence level. The error in the measured value is estimated with a sum of squares value of a 5\% systematic error due to the calibration of the equipment as well a statistical error given by SAODS9 in the flux measurement.\\

Annihilation channels  b$\overline{\textrm{b}}$, $ \tau ^+ \tau ^- $, $\mu^+ \mu^-$ and 2HDM+S were considered.  
\begin{figure}
	
	\includegraphics[width=0.48\linewidth]{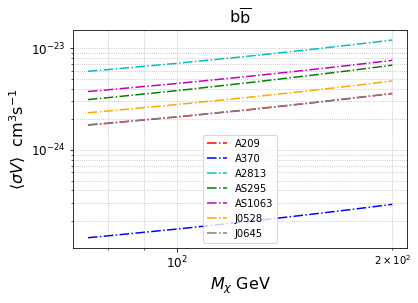} \hfill
	\label{fig:bb}
	\includegraphics[width=0.48\linewidth]{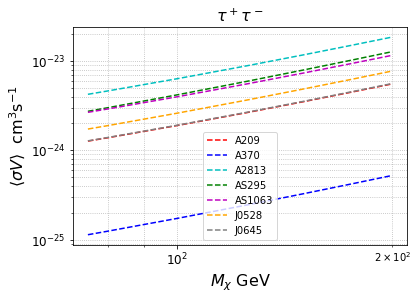}
	\label{fig:tau}

	\includegraphics[width=0.48\linewidth]{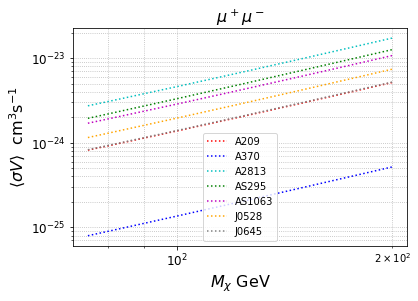} \hfill
	\label{fig:mu}
	\includegraphics[width=0.5\linewidth]{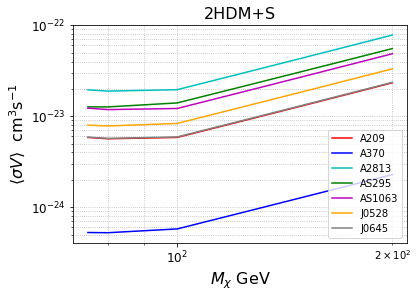}
	\label{fig:madala}
	
	\caption{Annihilation rates at a $2\sigma$ confidence level for the annihilation channels over the mass range 75-200 GeV}
	
\end{figure}

The cross section limits produced are above the thermal relic value, $<\sigma V> \approx10^{-26}$. Thus the dark matter model can not be ruled out as a candidate for all dark matter, as its present abundance may be less than what is required to agree with present cosmological constraints. \\ 	
In future work the radial surface brightness profiles will be fitted against predictions. It is expected that due to the high resolution of MeerKAT that this will be able to probe much lower cross sections. Clusters without identified halos can also be studied, as improved results may be attainable due to the clash in shape of a baryonic surface brightness profile and a dark matter surface brightness profile.

\ack{Acknowledgements}
NL acknowledges the financial assistance of the South African Radio Observatory (SARAO) towards this research (www.sarao.ac.za). G.B acknowledges support from a National Research Foundation of South Africa Thuthuka grant no. 117969.\\


	\bibliographystyle{unsrt} 
	
	{\bibliography{ref}}
\end{document}